\documentclass[aps,prl,twocolumn,superscriptaddress,amsmath,amssymb,floatfix]{revtex4-2}
\usepackage{amsthm,mathtools}
\usepackage{physics}
\usepackage{mleftright}
\mleftright
\usepackage{booktabs,array,graphicx}
\usepackage{xcolor}
\usepackage[colorlinks=true,linkcolor=blue!50!black,citecolor=blue!50!black,urlcolor=blue!50!black]{hyperref}
\hypersetup{pdfcreator={},pdfproducer={}}
\ifdefined\pdfinfoomitdate
  \pdftrailerid{}
\fi

\theoremstyle{definition}
\newtheorem{theorem}{Theorem}
\newtheorem{proposition}[theorem]{Proposition}

\newtheorem{corollary}[theorem]{Corollary}
\newtheorem{conjecture}[theorem]{Conjecture}
\newtheorem{definition}[theorem]{Definition}

\newcommand{\A}{\mathcal{A}}
\newcommand{\D}{\mathcal{D}}
\newcommand{\calJ}{\mathcal{J}}

\newcommand{\TL}{\mathrm{TL}}
\DeclareMathOperator{\sgn}{sgn}

\begin{document}

\title{Calculating correlation functions in integrable spin chains \\ from local conserved charges and currents}

\author{Kohei Fukai}
\email{kohei.fukai@riken.jp}
\affiliation{RIKEN iTHEMS, Wako, Saitama 351-0198, Japan}

\date{\today}

\begin{abstract}
We develop a simple algebraic method for calculating correlation functions in integrable spin chains.
We introduce current minors, antisymmetrized products of local conserved charge densities and their currents, and prove that their expectation values are determinants of density and current mean values.
Decomposing local operators into current minors, we derive factorized formulas for correlation functions in terms of current mean values, valid in any nondegenerate eigenstate.
Applying the method to the spin-$1/2$ XXZ chain and, more generally, to every representation of the Temperley--Lieb algebra, we prove previously conjectured factorized formulas for short-distance correlation functions.
\end{abstract}

\maketitle

\paragraph{Introduction.---}
Calculating correlation functions is a notoriously difficult problem even in integrable systems that are exactly solvable by the Bethe ansatz~\cite{Bethe1931,Baxter1982,KorepinBogoliubovIzergin1993}.
They have been studied by multiple integrals~\cite{JimboMikiMiwaNakayashiki1992,JimboMiwa1995,JimboMiwa1996,KitanineMailletTerras2000}, the reduced quantum Knizhnik--Zamolodchikov (qKZ) equations~\cite{BoosKorepinSmirnov2003,BJMST2006rqKZ}, and the hidden Grassmann structure~\cite{BJMST2007HGS1,BJMST2009HGS2,JMS2009HGS3}, with explicit results at short distances~\cite{SakaiShiroishiNishiyamaTakahashi2003,BoosShiroishiTakahashi2005,SatoShiroishiTakahashi2005,SatoShiroishiTakahashi2006,BoosGoehmannKlumperSuzuki2007,BoosDamerauGoehmann2008,TrippeGoehmannKlumper2010,SatoAufgebauerBoosGoehmannKlumperTakahashiTrippe2011,GoehmannKleinemuehlWeisse2021}.
In the spin-$1/2$ XXZ chain, the correlation functions were found to factorize: they are sums of products of the Taylor coefficients of a few state-dependent functions~\cite{BJMST2006Algebraic,Smirnov2021QuantumGroup}, and the same was conjectured for finite-volume excited states~\cite{Pozsgay2017Excited}.
For one of these functions, the coefficients are the mean values of the charge densities and generalized currents of the local conserved charges~\cite{BorsiPozsgayPristyak2020,FukaiKleinemuehlPozsgayVernier2024}.
The generalized currents describe the flow of one charge under the time evolution generated by another; they are central to generalized hydrodynamics~\cite{CastroAlvaredoDoyonYoshimura2016,BertiniColluraDeNardisFagotti2016}, and their mean values were obtained exactly from the Bethe roots~\cite{BorsiPozsgayPristyak2020,Pozsgay2020Currents,Pozsgay2020Algebraic,BorsiPristyakPozsgay2021}.

The resulting factorized formulas are simple, whereas their derivations are very involved, and some of the formulas are only conjectured~\cite{Pozsgay2017Excited,FukaiKleinemuehlPozsgayVernier2024}.
This suggests that they should be understood more simply, at the level of local operators.
The Hamiltonian of the XXZ chain is $H=\sum_je_j$, where the $e_j$ are the generators of the Temperley--Lieb (TL) algebra~\cite{TemperleyLieb1971}, and its charge densities and currents are expressed in terms of the $e_j$.
A first step is the identity that expresses $e_1e_2$ as a linear combination of these densities and currents~\cite[Eq.~(70)]{FukaiKleinemuehlPozsgayVernier2024}: $\ev{e_1e_2}$ is linear in their expectation values.
For other correlation functions, such as $\ev{e_1e_3}$ and $\ev{e_1e_4}$, formulas quadratic in the density and current mean values were found numerically~\cite[Eqs.~(71)--(73)]{FukaiKleinemuehlPozsgayVernier2024}.
Can these formulas be derived from identities among local operators, by elementary means?

In this Letter, we give such a derivation, by a method that applies to any translationally invariant integrable spin chain.
We introduce local operators called \emph{current minors}: alternating sums of ordered products of charge densities and currents.
We prove that the expectation value of a current minor in a nondegenerate simultaneous eigenstate of the charges is a determinant of current mean values~(Theorem~\ref{thm:minor}).
The proof is elementary: it uses only the continuity equations of the currents, and neither the Bethe ansatz nor the quantum inverse scattering method.
Once a local operator is decomposed into current minors, its expectation value follows immediately (Corollary~\ref{cor:bethe}).

For the TL algebra, a decomposition of a product of the $e_j$ into current minors is an identity in the algebra, so the resulting formula holds in every representation.
We give such decompositions and thereby prove Eqs.~(71)--(73) of Ref.~\cite{FukaiKleinemuehlPozsgayVernier2024} in any nondegenerate eigenstate, including finite-volume excited states.
We conjecture that every local TL operator admits a decomposition into current minors, so that every correlation function of the $e_j$ can be calculated by our method (Conjecture~\ref{conj:generation}).

\paragraph*{Charges and currents.---}
We first formulate the method for a general translationally invariant integrable spin chain of $L$ sites with periodic boundary conditions.
A characteristic feature of such a chain is the existence of an extensive number of local conserved charges:
\begin{equation}\label{eq:charges}
	Q_a=\sum_{x=1}^L q_a(x)\qquad(a\ge1),
\end{equation}
where the range of the density $q_a(x)$ grows with $a$, $Q_1=H$ is the Hamiltonian~\footnote{We follow the normalization of Ref.~\cite{FukaiKleinemuehlPozsgayVernier2024}, whose $Q_{a+1}$ and $J_{a+1,b+1}$ are our $Q_a$ and $J_{a,b}$.}, $q_a(x+1)=Uq_a(x)U^{-1}$, and $U$ is the translation operator.
The charges commute with each other,
\begin{equation}\label{eq:commuting}
	\comm{Q_a}{Q_b}=0.
\end{equation}

The building blocks of the method are the densities of the local conserved charges and their generalized currents.
Writing $\partial_bX=\comm{Q_b}{X}$, we define the generalized current $J_{a,b}(x)$, a local operator, by the continuity equation~\cite{FukaiKleinemuehlPozsgayVernier2024}
\begin{equation}\label{eq:continuity}
	\partial_bq_a(x)=J_{a,b}(x)-J_{a,b}(x+1),
\end{equation}
where $J_{a,b}(x+1)=UJ_{a,b}(x)U^{-1}$~\footnote{Each operator identity of this Letter involves operators of finite range; we take $L$ large compared with their range, so that none of them wraps around the chain.}.
We collect the densities and currents in the matrix $\calJ$:
\begin{equation}\label{eq:P}
	\calJ_{r0}(x)=q_{r+1}(x),\quad \calJ_{rc}(x)=J_{r+1,c}(x)\quad(c\ge1).
\end{equation}
Writing $\partial_0X=X-UXU^{-1}$ for the lattice difference, we have
\begin{equation}\label{eq:flat}
	\partial_c\calJ_{rb}(x)=\partial_b\calJ_{rc}(x).
\end{equation}
For $c=0$, this is Eq.~\eqref{eq:continuity}; for $b,c\ge1$, it follows from Eqs.~\eqref{eq:continuity} and~\eqref{eq:commuting} (Sec.~\ref{sm:flat}~\cite{SM}).

\paragraph*{Current minors.---}
The central objects of this Letter are minors of the matrix $\calJ$ of Eq.~\eqref{eq:P}.
\begin{definition}\label{def:minor}
Let $r_1,\ldots,r_n\ge0$, and let $c_1,\ldots,c_n\ge0$ be distinct.
The current minor is
\begin{equation}\label{eq:minor}
	\D_{r_1\cdots r_n;c_1\cdots c_n}(x)=\sum_{\sigma\in S_n}\sgn(\sigma)\prod_{i=1}^n\calJ_{r_ic_{\sigma(i)}}\bigl(x_i^\sigma\bigr),
\end{equation}
with factors multiplied from left to right in increasing $i$.
Here $x_i^\sigma=x+1$ if $c_{\sigma(j)}=0$ for some $j<i$, and $x_i^\sigma=x$ otherwise; that is, the factors to the right of the density are shifted by one.
For $n=0$, the minor is $1$.
\end{definition}
For example, $\D_{01;01}(x)=q_1(x)J_{2,1}(x+1)-J_{1,1}(x)q_2(x)$.

\paragraph*{Determinant formula.---}
We calculate the expectation values of the current minors in eigenstates of the charges.
We take the charges $Q_a$ and $U$ to be diagonalizable.
Let $\ket{E}$ be a simultaneous eigenstate of the local charges $Q_a$ and the translation operator $U$, and $\bra{E}$ the left eigenvector with the same eigenvalues, normalized by $\braket{E}=1$; we write $\ev{X}=\ev{X}{E}$.
Then $\ev{\partial_bX}=0$ for every operator $X$ and every $b\ge0$: with the eigenvalue $\epsilon_b$ of $Q_b$, $\ev{Q_bX-XQ_b}=(\epsilon_b-\epsilon_b)\ev{X}=0$, and similarly $\ev*{X-UXU^{-1}}=0$.
We call $\ket{E}$ \emph{nondegenerate} if the simultaneous eigenspace containing $\ket{E}$ is one-dimensional; such states are called singlets of the commuting set of charges in Ref.~\cite{FukaiKleinemuehlPozsgayVernier2024}.

The results below depend on the eigenstate only through the density and current mean values
\begin{equation}\label{eq:psi}
	\psi_{rc}=\ev{\calJ_{rc}(x)},
\end{equation}
which do not depend on $x$.
\begin{theorem}\label{thm:minor}
If $\ket{E}$ is nondegenerate, then
\begin{equation}\label{eq:det}
	\ev{\D_{r_1\cdots r_n;c_1\cdots c_n}(x)}=\det\bigl[\psi_{r_ic_j}\bigr]_{i,j=1}^n .
\end{equation}
\end{theorem}

\begin{proof}
We first remove the translation from Eq.~\eqref{eq:minor}.
A term whose density $\calJ_{r_k0}(x)$ is in row $k$ has the form $A\,\calJ_{r_k0}(x)\,UBU^{-1}$, where $A$ and $B$ are the products of the factors to the left and right of it, all taken at $x$.
With $U\ket{E}=e^{ip_E}\ket{E}$ and $V_0=e^{-ip_E}U$, we have $V_0^{-1}\ket{E}=\ket{E}$, so $\ev{A\,\calJ_{r_k0}(x)\,UBU^{-1}}=\ev{A\,\calJ_{r_k0}(x)V_0\,B}$.
Let $\{\ket{\nu}\}$ be a basis of simultaneous eigenvectors containing $\ket{E}$, with dual basis $\{\bra{\nu}\}$, and write $X_{\nu\nu'}=\mel{\nu}{X}{\nu'}$.
Inserting $\sum_\nu\dyad{\nu}$ between consecutive factors, the factor in row $i$ always lies between $\bra{\nu_{i-1}}$ and $\ket{\nu_i}$, so
\[
	\ev{\D_{r_1\cdots r_n;c_1\cdots c_n}(x)}=\sum_{\nu_1,\ldots,\nu_{n-1}}\det\bigl[T_{ic_j}\bigr]_{i,j=1}^n,
\]
where $\nu_0=\nu_n=E$ and, for every $b\ge0$,
\[
	T_{ib}=\begin{cases}(\calJ_{r_ib}(x))_{\nu_{i-1}\nu_i},&b\ge1,\\[2pt](\calJ_{r_i0}(x)V_0)_{\nu_{i-1}\nu_i},&b=0.\end{cases}
\]

Next, Eq.~\eqref{eq:flat} makes $T_{ib}$ proportional to eigenvalue differences.
Let $\epsilon_b(\nu)$ be the eigenvalue of $Q_b$ ($b\ge1$) and $\epsilon_0(\nu)=-e^{i(p_\nu-p_E)}$ that of $-V_0$, and put $\Delta_{ib}=\epsilon_b(\nu_{i-1})-\epsilon_b(\nu_i)$.
Between $\bra{\nu_{i-1}}$ and $\ket{\nu_i}$, the commutator $\partial_b=\comm{Q_b}{\cdot}$ ($b\ge1$) multiplies matrix elements by $\Delta_{ib}$, and so does $(\partial_0\,\cdot\,)V_0=\comm{-V_0}{\cdot}$.
For $b,c\ge1$, the matrix element of Eq.~\eqref{eq:flat} is therefore $\Delta_{ic}T_{ib}=\Delta_{ib}T_{ic}$.
For $c=0$, Eq.~\eqref{eq:flat} multiplied by $V_0$ from the right reads $(\partial_0\calJ_{r_ib}(x))V_0=\partial_b(\calJ_{r_i0}(x)V_0)$, since $\comm{Q_b}{V_0}=0$, and its matrix element is $\Delta_{i0}T_{ib}=\Delta_{ib}T_{i0}$.
Altogether,
\begin{equation}\label{eq:flat-T}
	\Delta_{ic}\,T_{ib}=\Delta_{ib}\,T_{ic}\qquad(b,c\ge0).
\end{equation}
Hence, if $\Delta_{ib}\ne0$ for some $b$, there is $\varphi_i$ with
\begin{equation}\label{eq:gradient}
	T_{ib}=\varphi_i\,\Delta_{ib}\qquad(b\ge0).
\end{equation}

Finally, the rows of $T$ at the steps that leave an eigenspace are linearly dependent.
Let $I$ be the set of $i$ for which $\Delta_{ib}\ne0$ for some $b$.
If $I\ne\emptyset$, either some $\varphi_i=0$, or
\[
	\sum_{i\in I}\frac{T_{ic_j}}{\varphi_i}=\sum_{i=1}^n\Delta_{ic_j}=\epsilon_{c_j}(\nu_0)-\epsilon_{c_j}(\nu_n)=0;
\]
in both cases the determinant vanishes.
If $I=\emptyset$, then $\nu_i=E$ for all $i$ by nondegeneracy, and $T_{ib}=\psi_{r_ib}$.
\end{proof}

For the minors $\D_{rs;0t}(x)=\calJ_{r0}(x)\calJ_{st}(x+1)-\calJ_{rt}(x)\calJ_{s0}(x)$ ($t\ge1$), Theorem~\ref{thm:minor} gives $\ev{\D_{rs;0t}(x)}=\psi_{r0}\psi_{st}-\psi_{rt}\psi_{s0}$, which was proved by a different method in Ref.~\cite{PozsgayJiangTakacs2020}.
Theorem~\ref{thm:minor} extends this formula to current minors of any order.

We define the subspace of lattice differences and commutators with the charges,
\begin{equation}\label{eq:dA}
	\partial\A=\Bigl\{\sum_{b\ge0}\partial_bA_b\Bigm|A_b\in\A\Bigr\},
\end{equation}
where $\A$ is the space of operators of the chain, and write $X\equiv Y\pmod{\partial\A}$ if $X-Y\in\partial\A$; since $\ev{\partial_bX}=0$, the elements of $\partial\A$ have zero expectation value.
A current minor $\D_{R;C}(x)$, where $R=r_1\cdots r_n$ and $C=c_1\cdots c_n$, does not depend on $x$ modulo $\partial\A$, since $\D_{R;C}(x)-\D_{R;C}(x+1)=\partial_0\D_{R;C}(x)\in\partial\A$; we omit its argument when we work modulo $\partial\A$.

\begin{corollary}\label{cor:bethe}
Let $O\in\A$ satisfy
\begin{equation}\label{eq:decomposition}
	O\equiv\sum_{I}c_I\,\D_{R_I;C_I}\pmod{\partial\A}
\end{equation}
with scalar coefficients $c_I$.
If $\ket{E}$ is nondegenerate, then
\begin{equation}\label{eq:bethe-general}
	\ev{O}=\sum_Ic_I\det\psi_{R_I,C_I}.
\end{equation}
\end{corollary}

\paragraph*{Temperley--Lieb algebra.---}
We now apply these results to the periodic TL algebra $\TL$ with parameter $d$.
It is generated by $e_1,\ldots,e_L$ with the relations
\begin{equation}\label{eq:TL}
	e_j^2=d\,e_j,\quad e_je_{j\pm1}e_j=e_j,
\end{equation}
and $\comm{e_j}{e_l}=0$ for $l\ne j,j\pm1$, where the indices are taken modulo $L$~\cite{Jones1983,MartinSaleur1993,Nichols2006,FukaiKleinemuehlPozsgayVernier2024}.
The Hamiltonian is defined abstractly as
\begin{equation}\label{eq:hamiltonian}
	H=\sum_{j=1}^L e_j.
\end{equation}
The translation acts on $\TL$ by $Ue_jU^{-1}=e_{j+1}$~\footnote{In the Potts chain, the translation by one site shifts the TL indices by two, and the results hold for expectation values averaged over two neighboring TL indices; see Ref.~\cite{FukaiKleinemuehlPozsgayVernier2024}.}, and the results above apply in every representation, such as the Potts chain~\cite{Martin1991} and the golden chain~\cite{Feiguin2007}.
The spin-$1/2$ XXZ chain is the representation on $(\mathbb C^2)^{\otimes L}$ with $d=q+q^{-1}$ and
\begin{equation}\label{eq:xxz}
\begin{aligned}
	e_j=\frac12\Bigl[&-\sigma^x_j\sigma^x_{j+1}-\sigma^y_j\sigma^y_{j+1}+\frac d2\bigl(1-\sigma^z_j\sigma^z_{j+1}\bigr)\\
	&+\frac{q^{-1}-q}{2}\bigl(\sigma^z_j-\sigma^z_{j+1}\bigr)\Bigr].
\end{aligned}
\end{equation}

The TL Hamiltonian~\eqref{eq:hamiltonian} has an extensive number of local conserved charges~\cite{NienhuisHuijgen2021,FukaiKleinemuehlPozsgayVernier2024}.
The first charge densities are $q_1(x)=e_x$ and $q_2(x)=e_xe_{x+1}-e_{x+1}e_x$, and in general $q_a(x)$ is a polynomial in $e_x,\ldots,e_{x+a-1}$; in the XXZ chain, it acts on the $a+1$ sites $x,\ldots,x+a$.
The charges are generated formally by the boost operator $B=\sum_x x e_x$ of the infinite chain through $Q_{a+1}=\comm{Q_a}{B}$~\cite{Tetelman1982}; explicit densities in terms of the $e_j$ are given in Refs.~\cite{NienhuisHuijgen2021,FukaiKleinemuehlPozsgayVernier2024} and in Sec.~\ref{sm:densities}~\cite{SM}.

In the periodic XXZ chain, the mean values $\psi_{rc}$ are known in closed form~\cite{BorsiPozsgayPristyak2020,Pozsgay2020Currents,FukaiKleinemuehlPozsgayVernier2024}.
For a regular Bethe state with $M$ magnons, Bethe roots $\lambda_1,\ldots,\lambda_M$, and an invertible Gaudin matrix $\Gamma$,
\begin{equation}\label{eq:psi-gaudin}
	\psi_{rc}=\sum_{j,k=1}^M h_{r+1}(\lambda_j)\,(\Gamma^{-1})_{jk}\,h_{c+1}(\lambda_k),
\end{equation}
where $h_a(\lambda)$ is the one-particle eigenvalue of $Q_a$; for $c=0$, this is $\psi_{r0}=L^{-1}\sum_jh_{r+1}(\lambda_j)$.
The explicit forms of $h_a$ and $\Gamma$ are given in Sec.~\ref{sm:bethe}~\cite{SM}.
In several other representations, the mean values $\psi_{rc}$ were found numerically to coincide with those of the XXZ chain with the same $d$~\cite{FukaiKleinemuehlPozsgayVernier2024}.

The mean values $\psi_{rc}$ also enter the factorized formulas of the XXZ chain~\cite{BorsiPozsgayPristyak2020,BorsiPristyakPozsgay2021,FukaiKleinemuehlPozsgayVernier2024}.
There, the expectation values of local operators in the TL algebra (equivalently, $U_q(\mathfrak{sl}_2)$-invariant ones) are polynomials in the Taylor coefficients of a single function $\omega(\lambda,\mu)$, with coefficients that depend only on $q$~\cite{BJMST2006Algebraic,JMS2009HGS3,Smirnov2021QuantumGroup}, and the state enters only through $\omega$~\cite{BJMST2006rqKZ,BoosDamerauGoehmann2008,TrippeGoehmannKlumper2010,Pozsgay2017Excited}.
In a regular Bethe state, $\omega$ is, up to a change of variables and an explicit additive term, the generating function of the current mean values $\psi_{rc}$~\cite{BorsiPozsgayPristyak2020}; see Ref.~\cite[Eqs.~(55)--(59)]{FukaiKleinemuehlPozsgayVernier2024} for the precise relation.

\paragraph*{Short-distance correlation functions.---}
We now prove the previously conjectured formulas, Eqs.~(71)--(73) of Ref.~\cite{FukaiKleinemuehlPozsgayVernier2024}, by Corollary~\ref{cor:bethe}, giving the decompositions explicitly.
We define $\partial\TL$ by Eq.~\eqref{eq:dA} with $\A=\TL$.
\begin{proposition}\label{prop:local}
Modulo $\partial\TL$,
\begin{equation}\label{eq:headline}
\begin{aligned}
e_1e_3\equiv{}& \frac{1}{12d(d^2-1)}
  \Bigl[12(d^2-4)\calJ_{00}+24\calJ_{02}\\
  &+(d^2-40)\calJ_{11}+2\calJ_{13}-3\calJ_{22}\\
  &-d(d^2-28)\D_{01;01}+3d\,\D_{02;02}\\
  &-6d\,\D_{12;01}-2d\,\D_{01;03}\Bigr].
\end{aligned}
\end{equation}
Analogous identities hold for $e_1e_2e_3+e_3e_2e_1$ and $e_1e_4$ (Eqs.~\eqref{eq:id012} and~\eqref{eq:id03}~\cite{SM}).
\end{proposition}

By Theorem~\ref{thm:minor} and the symmetry $\psi_{rc}=\psi_{cr}$~\cite{FukaiKleinemuehlPozsgayVernier2024,BorsiPozsgayPristyak2020,Pozsgay2020Algebraic} (Sec.~\ref{sm:flat}~\cite{SM}), Eq.~\eqref{eq:headline} gives
\begin{equation}\label{eq:e13-psi}
\begin{aligned}
\ev{e_1e_3}={}&\frac{1}{12(d^2-1)}\Bigl\{\frac1d\bigl[12(d^2-4)\psi_{00}+24\psi_{02}\\
&+(d^2-40)\psi_{11}+2\psi_{13}-3\psi_{22}\bigr]\\
&-(d^2-28)(\psi_{00}\psi_{11}-\psi_{01}^2)+3(\psi_{00}\psi_{22}-\psi_{02}^2)\\
&-6(\psi_{01}\psi_{12}-\psi_{02}\psi_{11})-2(\psi_{00}\psi_{13}-\psi_{01}\psi_{03})\Bigr\},
\end{aligned}
\end{equation}
where each quadratic term comes from one current minor in Eq.~\eqref{eq:headline}.
This is Eq.~(71) of Ref.~\cite{FukaiKleinemuehlPozsgayVernier2024}.
The other two identities prove Eqs.~(72) and~(73) of Ref.~\cite{FukaiKleinemuehlPozsgayVernier2024} in the same way.
We also give a decomposition of $e_1e_5$ with current minors up to order three (Sec.~\ref{sm:short}~\cite{SM}).

\begin{conjecture}\label{conj:generation}
For every $O\in\TL$, there are rational functions $c_I(d)$ of $d$ such that
\begin{equation}\label{eq:generation}
	O\equiv\sum_Ic_I(d)\,\D_{R_I;C_I}\pmod{\partial\TL}.
\end{equation}
\end{conjecture}
By Corollary~\ref{cor:bethe}, it implies Conjecture~2 of Ref.~\cite{FukaiKleinemuehlPozsgayVernier2024} in nondegenerate eigenstates.
We expect that Conjecture~\ref{conj:generation} can be proved by a method related to the reduced qKZ equations~\cite{BJMST2006rqKZ,AufgebauerKlumper2012}.

\paragraph*{Outlook.---}
Theorem~\ref{thm:minor} and Corollary~\ref{cor:bethe} hold in any translationally invariant integrable spin chain.
There, every local operator decomposed into current minors has a factorized expectation value, and the current mean values of the model then give its correlation functions explicitly, in the ground state and in excited states.
The decompositions need explicit charge densities, which are known for several integrable chains~\cite{Anshelevich1980,GrabowskiMathieu1994,GrabowskiMathieu1995,NozawaFukai2020,Nozawa2022Thesis,YamadaFukai2023,FukaiYamada2026,Fukai2023Hubbard,Fukai2024Thesis,Fukai2024HubbardCompleteness,FukaiPozsgayVona2026II,SanataniShiraishi2026,IwasakiKatsura2026}.
The method applies immediately to the XYZ chain, where the current mean values have also been obtained~\cite{PristyakPozsgay2023}.

The hidden Grassmann structure gives the factorized formulas of the infinite XXZ chain through a fermionic basis of local operators, whose expectation values are expressed in terms of the function $\omega$~\cite{BJMST2007HGS1,BJMST2009HGS2,JMS2009HGS3,Smirnov2021QuantumGroup}.
Since $\omega$ generates the mean values $\psi_{rc}$, we expect a direct relation between the fermionic basis and current minors, which remains to be clarified.

An open problem is how to calculate the correlation functions of local operators that cannot be decomposed into current minors as in Corollary~\ref{cor:bethe}.
Local operators of the XXZ chain that are not generated by the $e_j$, such as $\sigma^z_1\sigma^z_r$, lie outside the TL algebra, and their expectation values also involve a second function $\tilde\omega$~\cite{BJMST2006rqKZ,BJMST2006Algebraic}.
In the ground state of the infinite chain, $\omega$ and $\tilde\omega$ are, up to a common factor, the derivatives of the logarithm of a single function with respect to the spectral parameter and to the anisotropy~\cite[Eq.~(3.13)]{BJMST2006Algebraic}.
We expect the anisotropy derivatives of the charge densities and currents to supply the operators missing from the TL algebra, and current minors extended by them to describe every local correlation function of the XXZ chain.

\paragraph{Acknowledgments.---}
The author used models from Anthropic (Claude Fable~5.2 and Claude Opus~5.5) and \mbox{OpenAI} (GPT-6~Astra and GPT-6.1~Sol) for assistance with the proof of Theorem~\ref{thm:minor}, the computer-algebra code, and the writing of the manuscript.
The author checked all derivations and computations and takes full responsibility for the content.
The author was supported by JSPS KAKENHI Grant-in-Aid for Early-Career Scientists (Grant No.~JP26K17045), JSPS KAKENHI Grant-in-Aid for JSPS Fellows (Grant No.~JP26KJ0404), and the JSPS Bilateral Program with MTA (Project No.~120263802).

\paragraph*{Data availability.---}
The code and data that support Proposition~\ref{prop:local} and Conjecture~\ref{conj:generation} for the operators generated by $e_1,\ldots,e_4$, including the operators $A_b$ that give the terms in $\partial\TL$, are openly available~\cite{FukaiZenodo2026}.

\bibliography{references}

\clearpage
\onecolumngrid
\begin{center}
\textbf{\large Supplemental Material for\\[2pt] ``Calculating correlation functions in integrable spin chains from local conserved charges and currents''}
\end{center}
\twocolumngrid

\setcounter{secnumdepth}{1}
\setcounter{section}{0}
\setcounter{equation}{0}
\setcounter{table}{0}
\setcounter{theorem}{0}
\renewcommand{\thesection}{S\arabic{section}}
\renewcommand{\theequation}{S\arabic{equation}}
\renewcommand{\theHequation}{S\arabic{equation}}
\renewcommand{\thetable}{S\arabic{table}}
\renewcommand{\thetheorem}{S\arabic{theorem}}

\section{Densities and currents}\label{sm:densities}

In this section, we give the explicit densities and currents.
We take $q_1(x)=e_x$, and for $a\ge1$ obtain $q_{a+1}(x)$ by replacing each reduced word in $J_{1,a}(x)$ by its translate with smallest index $x$~\cite{FukaiKleinemuehlPozsgayVernier2024}.
The first densities are
\begin{equation}\label{eq:q34}
\begin{aligned}
	q_2(x)&=e_xe_{x+1}-e_{x+1}e_x,\\
	q_3(x)&=2(e_{x+2}e_{x+1}e_x-e_{x+1}e_xe_{x+2}\\
	&\quad-e_xe_{x+2}e_{x+1}+e_xe_{x+1}e_{x+2})\\
	&\quad+d(e_{x+1}e_x+e_xe_{x+1}).
\end{aligned}
\end{equation}
The currents $J_{a,b}(x)$ then follow from Eq.~\eqref{eq:continuity}~\cite{FukaiKleinemuehlPozsgayVernier2024}, where we fix the scalar by requiring that $J_{a,b}(x)$ has no term proportional to the identity.

The simplest minor of currents only is $\D_{01;12}(x)=J_{1,1}(x)J_{2,2}(x)-J_{1,2}(x)J_{2,1}(x)$, where $J_{1,2}(x)$ and $J_{2,2}(x)$ have $10$ and $22$ reduced words.
At $x=3$, where only $e_1,\ldots,e_5$ occur, it reads
\begin{equation}\label{eq:D0112}
\begin{aligned}
	\D_{01;12}(3)={}&-e_1e_2-e_2e_1+2e_2e_3+2e_2e_4-e_3e_2\\
	&-e_3e_4+2e_4e_3-e_1e_2e_3e_4+e_1e_2e_4e_3\\
	&+2e_1e_3e_2e_4+2e_2e_1e_3e_2+e_2e_1e_3e_4\\
	&+e_2e_1e_4e_3+e_2e_3e_4e_5-e_2e_3e_5e_4\\
	&-e_2e_4e_3e_5+e_2e_5e_4e_3-e_3e_2e_4e_3\\
	&-2e_2e_1e_3e_2e_4e_3-2e_3e_2e_1e_4e_3e_2\\
	&+d\bigl(-2e_2+3e_3-e_4+2e_1e_2e_3+e_1e_2e_4\\
	&\quad +2e_1e_3e_4-2e_2e_1e_3+e_2e_1e_4\\
	&\quad -e_2e_3e_4-e_2e_4e_5+e_2e_5e_4+2e_3e_2e_4\\
	&\quad -e_3e_4e_5+e_3e_5e_4-e_1e_3e_2e_4e_3\\
	&\quad -2e_2e_1e_3e_2e_4-e_3e_2e_1e_4e_3\\
	&\quad +e_3e_2e_4e_3e_5-e_3e_2e_5e_4e_3\bigr)\\
	&+d^2\bigl(2e_2e_3-2e_3e_4\bigr) .
\end{aligned}
\end{equation}
By Theorem~\ref{thm:minor}, its expectation value in a nondegenerate state is $\psi_{01}\psi_{12}-\psi_{02}\psi_{11}$.

\section{Proof of \texorpdfstring{Eq.~(\ref{eq:flat})}{Eq. (5)}}\label{sm:flat}

Equation~\eqref{eq:flat} is symmetric in $b$ and $c$, and for $c=0$ it is Eq.~\eqref{eq:continuity}.
Let $b,c\ge1$, and set
\begin{equation}\label{eq:flat-defect}
	K(x)=\partial_c\calJ_{rb}(x)-\partial_b\calJ_{rc}(x).
\end{equation}
Since the charges are translation invariant, $\partial_0$ commutes with $\partial_b$ and $\partial_c$, and Eqs.~\eqref{eq:continuity} and~\eqref{eq:commuting} give
\begin{equation}\label{eq:flat-proof}
\begin{aligned}
	K(x)-K(x+1)&=\partial_c\partial_b\calJ_{r0}(x)-\partial_b\partial_c\calJ_{r0}(x)\\
	&=\comm{\comm{Q_c}{Q_b}}{\calJ_{r0}(x)}=0 .
\end{aligned}
\end{equation}
We show that this forces $K(x)=0$.
In a spin chain, $K(x)$ acts on finitely many sites and commutes with $U$, so it is a scalar; it is a sum of commutators and therefore traceless, so $K(x)=0$.
In the TL algebra, we use reduced words in place of the trace.
We say that $X$ is supported on $[j,k]$ if it is written with the generators $e_j,\ldots,e_k$ only.
The currents $\calJ_{rb}(x)$ and $\calJ_{rc}(x)$ are supported on finite intervals, and so is $K(x)$, since in $\partial_bX=\sum_y\comm{q_b(y)}{X}$ only the densities $q_b(y)$ near the support of $X$ fail to commute with $X$.
Let $X$ be supported on $[j,k]$, and assume $UXU^{-1}=X$, so that $U^MXU^{-M}=X$ for every $M$.
Expanding $X$ in reduced words $w$ in $e_j,\ldots,e_k$, we have, for $M>k-j$,
\begin{align}
	X&=c_1+\sum_{w\ne1}c_w\,w, \notag\\
	X=U^MXU^{-M}&=c_1+\sum_{w\ne1}c_w\,U^MwU^{-M},\label{eq:local-scalar}
\end{align}
where $U^MwU^{-M}$ is a reduced word in $e_{j+M},\ldots,e_{k+M}$.
For $L$ large, no $U^MwU^{-M}$ with $w\ne1$ occurs among the words $w$, and reduced words are linearly independent~\cite{Jones1983,Martin1991}.
Comparing the two lines gives $c_w=0$ for $w\ne1$, so $X=c_1$ is a scalar; in particular $K(x)$ is a scalar.

Finally, the relations~\eqref{eq:TL} never remove all generators from a product of words that each contain a generator, and $q_b(y)$ has no term proportional to the identity.
Hence no commutator $\partial_bX=\sum_y\comm{q_b(y)}{X}$ with $b\ge1$ has a term proportional to the identity, so neither has $K(x)$, and $K(x)=0$.

In the TL algebra, Eq.~\eqref{eq:flat} implies the symmetry $\psi_{rc}=\psi_{cr}$, which was observed in Ref.~\cite[Eq.~(64)]{FukaiKleinemuehlPozsgayVernier2024}; see also Refs.~\cite{BorsiPozsgayPristyak2020,Pozsgay2020Algebraic}.
In our normalization, $\calJ_{0r}(x)-q_{r+1}(x)=\partial_0g_r(x)$ for a local $g_r(x)$.
Replacing the density $q_{r+1}(x)$ by $\calJ_{0r}(x)$ leaves the total charge fixed and changes its currents to $\calJ'_{rc}(x)=\calJ_{rc}(x)+\partial_cg_r(x)$.
For $r,c\ge1$, $\partial_0(\calJ'_{rc}(x)-\calJ'_{cr}(x))=\partial_c\calJ_{0r}(x)-\partial_r\calJ_{0c}(x)=0$ by Eq.~\eqref{eq:flat}, so $\calJ'_{rc}(x)-\calJ'_{cr}(x)$ is a scalar by Eq.~\eqref{eq:local-scalar}, and it vanishes because neither current has a term proportional to the identity.
Hence $\calJ_{rc}\equiv\calJ_{cr}\pmod{\partial\TL}$; for $r=0$ or $c=0$ this follows from $\calJ_{0r}(x)-q_{r+1}(x)=\partial_0g_r(x)$.
A sufficient finite-size condition is $L\ge r+c+2$ and a simultaneous eigenstate of $Q_1,\ldots,Q_{r+c-1}$ and $U$; for $r+c\le1$ only translation invariance is needed.
The entry $\psi_{\alpha,\beta}$ of Ref.~\cite{FukaiKleinemuehlPozsgayVernier2024} is our $\psi_{\beta\alpha}$.

\section{Current mean values in the XXZ chain}\label{sm:bethe}

In this section, we give the functions $h_a$ and the matrix $\Gamma$ of Eq.~\eqref{eq:psi-gaudin} in the normalization of the main text.
For the periodic XXZ chain with $d=2\cos\gamma$, the Bethe roots $\lambda_1,\ldots,\lambda_M$ of a state with $M$ magnons and zero twist satisfy, for $j=1,\ldots,M$,
\begin{equation}\label{eq:bae}
	\Bigl(\frac{\sinh(\lambda_j+i\gamma/2)}{\sinh(\lambda_j-i\gamma/2)}\Bigr)^L
	\prod_{k\ne j}\frac{\sinh(\lambda_j-\lambda_k-i\gamma)}{\sinh(\lambda_j-\lambda_k+i\gamma)}=1.
\end{equation}
The one-particle eigenvalue $h_a(\lambda)$ of $Q_a$ is given by
\[
\begin{aligned}
	h_1(\lambda)&=\frac{\sin^2\gamma}{\sinh(\lambda+i\gamma/2)\sinh(\lambda-i\gamma/2)},\\
	h_{a+1}(\lambda)&=-i\sin\gamma\,\partial_\lambda h_a(\lambda).
\end{aligned}
\]
The matrix $\Gamma$, which is $\sin\gamma$ times the Gaudin matrix of Ref.~\cite[Eq.~(62)]{FukaiKleinemuehlPozsgayVernier2024}, is
\begin{equation}\label{eq:gaudin}
\begin{aligned}
	\Gamma_{jk}&=Lh_1(\lambda_j)\delta_{jk}+\mathcal K(\lambda_j-\lambda_k)-\delta_{jk}\sum_{\ell}\mathcal K(\lambda_j-\lambda_\ell),\\
	\mathcal K(\lambda)&=\frac{\sin\gamma\,\sin2\gamma}{\sinh(\lambda+i\gamma)\sinh(\lambda-i\gamma)}.
\end{aligned}
\end{equation}
Equation~\eqref{eq:psi-gaudin} holds for $r+c\le L-2$.
For $c=0$, Eq.~\eqref{eq:psi-gaudin} reduces to $\psi_{r0}=L^{-1}\sum_jh_{r+1}(\lambda_j)$ because $\sum_k\Gamma_{jk}=Lh_1(\lambda_j)$.

\section{Decompositions}\label{sm:short}

In this section, we give the identities that complete Proposition~\ref{prop:local}, prove the proposition, and give the decomposition of $e_1e_5$.
Modulo $\partial\TL$,
\begin{align}
&e_1e_2e_3+e_3e_2e_1\equiv \frac{1}{6d(d^2-1)}
  \Bigl[36d\,\calJ_{00}-24d\,\calJ_{02}\notag\\
  &\quad+(5d^3+34d)\calJ_{11}-2d\,\calJ_{13}+3d\,\calJ_{22}\notag\\
  &\quad-(11d^2+16)\D_{01;01}-3\D_{02;02}+6\D_{12;01}+2\D_{01;03}\Bigr],
  \label{eq:id012}\\[0.4em]
&e_1e_4\equiv \frac{1}{144d(d^2-1)(d^2-2)}
  \Bigl[288(d^2-4)\calJ_{00}\notag\\
  &\quad-12(11d^2-92)\calJ_{02}+4(d^4+40d^2-392)\calJ_{11}\notag\\
  &\quad+4(d^2+56)\calJ_{13}-6(d^2+44)\calJ_{22}\notag\\
  &\quad-12\calJ_{04}+6\calJ_{24}-8\calJ_{33}
  +\textstyle\sum_{r,s,c,t} c_{rs;ct}\,\D_{rs;ct}\Bigr],
  \label{eq:id03}
\end{align}
where the sum runs over the sixteen entries of Table~\ref{tab:minors-e03}.
To prove Eqs.~\eqref{eq:headline}, \eqref{eq:id012}, and~\eqref{eq:id03}, clear the denominator and subtract $\sum_{b\ge1}\partial_bA_b$, with explicit $A_b\in\TL$ obtained by computer algebra and given in Ref.~\cite{FukaiZenodo2026}.
For every reduced word, the coefficients of its translates in the remainder sum to zero.
Hence the remainder is a lattice difference: replacing each word by its translate with smallest index $1$ changes it by a lattice difference, and leaves the sums above, which vanish.

\begin{table}[b]
\caption{Coefficients $c_{rs;ct}$ in Eq.~\eqref{eq:id03}.}
\label{tab:minors-e03}
\begin{tabular}{@{}ll@{\qquad}ll@{}}
\toprule
minor & coefficient & minor & coefficient\\
\midrule
$\mathcal D_{01;01}$ & $- 8 d \left(d^{2} - 10\right) \left(5 d^{2} + 22\right)$ & $\mathcal D_{03;03}$ & $32 d$\\
$\mathcal D_{01;03}$ & $- 64 d \left(d^{2} + 8\right)$ & $\mathcal D_{12;12}$ & $6 d \left(5 d^{2} + 22\right)$\\
$\mathcal D_{02;02}$ & $96 d \left(d^{2} + 8\right)$ & $\mathcal D_{12;23}$ & $- 12 d$\\
$\mathcal D_{12;01}$ & $- d \left(7 d^{4} + 226 d^{2} + 928\right)$ & $\mathcal D_{13;13}$ & $8 d$\\
$\mathcal D_{14;01}$ & $d \left(5 d^{2} + 22\right)$ & $\mathcal D_{12;14}$ & $- 6 d$\\
$\mathcal D_{13;02}$ & $- 2 d \left(5 d^{2} + 22\right)$ & $\mathcal D_{23;03}$ & $- 4 d$\\
$\mathcal D_{23;01}$ & $- 96 d$ & $\mathcal D_{24;02}$ & $3 d$\\
$\mathcal D_{02;04}$ & $- 24 d$ & $\mathcal D_{34;01}$ & $- 2 d$\\
\bottomrule
\end{tabular}
\end{table}

The word $e_1e_5$ has the decomposition
\begin{equation}\label{eq:e15}
\begin{aligned}
 &34560\,d\,(d^2-1)^2(d^2-2)(d^4-3d^2+1)\,e_1e_5\\
 &\qquad\equiv\sum_Ic_I\,\D_{R_I;C_I}\pmod{\partial\TL},
\end{aligned}
\end{equation}
where the sum runs over the $73$ entries of Table~\ref{tab:minors-e15}: $11$ entries $\D_{r;c}=\calJ_{rc}$, $39$ minors of order two, and $23$ of order three.
Equation~\eqref{eq:e15} is proved as above, by subtracting $\sum_{b=1}^{4}\partial_bA_b$ with the explicit $A_b$ of Ref.~\cite{FukaiZenodo2026}.
By Corollary~\ref{cor:bethe}, $\ev{e_1e_5}$ is given by Eq.~\eqref{eq:e15} with each $\D_{R_I;C_I}$ replaced by $\det\psi_{R_I,C_I}$.
Its XXZ polynomial agrees with the fermionic expansion obtained by the procedure of Ref.~\cite[Sec.~4]{Smirnov2021QuantumGroup}.
At $d=2$, the corresponding dimer correlation in the ground state of the infinite antiferromagnetic XXX chain was evaluated in Ref.~\cite[Eq.~(3.15)]{SatoShiroishiTakahashi2006}.

\begin{table*}[b]
\caption{Coefficients $c_I$ of the current minors in Eq.~\eqref{eq:e15}.}
\label{tab:minors-e15}
\scriptsize
\resizebox{\textwidth}{!}{\begin{tabular}{@{}ll@{\quad}ll@{}}
\toprule
minor & coefficient & minor & coefficient\\
\midrule
$\mathcal D_{0;0}$ & $34560(3d^{6}-17d^{4}+22d^{2}-8)$ & $\mathcal D_{12;25}$ & $-6d(d^{8}-71d^{6}-89d^{4}-86d^{2}+56)$\\
$\mathcal D_{0;2}$ & $-2880(d^{8}+34d^{6}-283d^{4}+376d^{2}-128)$ & $\mathcal D_{12;34}$ & $10d(d^{8}-47d^{6}-329d^{4}+70d^{2}+32)$\\
$\mathcal D_{0;4}$ & $2880(d^{6}-10d^{4}+13d^{2}-4)$ & $\mathcal D_{12;45}$ & $-6d(d^{6}+5d^{4}+3d^{2}-2)$\\
$\mathcal D_{1;1}$ & $48(9d^{10}+77d^{8}+2118d^{6}-20868d^{4}+28648d^{2}-9984)$ & $\mathcal D_{13;13}$ & $8d(d^{10}+117d^{8}+2379d^{6}+8814d^{4}-2456d^{2}-1232)$\\
$\mathcal D_{1;3}$ & $48(5d^{8}-303d^{6}+5642d^{4}-7872d^{2}+2528)$ & $\mathcal D_{13;15}$ & $4d(d^{8}-47d^{6}-329d^{4}+70d^{2}+32)$\\
$\mathcal D_{1;5}$ & $48(d^{6}-39d^{4}+54d^{2}-16)$ & $\mathcal D_{13;24}$ & $10d(d^{8}-71d^{6}-89d^{4}-86d^{2}+56)$\\
$\mathcal D_{2;2}$ & $-60(5d^{8}-143d^{6}+4362d^{4}-6272d^{2}+2048)$ & $\mathcal D_{13;35}$ & $8d(d^{6}+5d^{4}+3d^{2}-2)$\\
$\mathcal D_{2;4}$ & $120(5d^{6}+141d^{4}-210d^{2}+64)$ & $\mathcal D_{14;14}$ & $-5d(d^{8}-47d^{6}-329d^{4}+70d^{2}+32)$\\
$\mathcal D_{3;3}$ & $-960(d^{6}+17d^{4}-26d^{2}+8)$ & $\mathcal D_{14;34}$ & $-10d(d^{6}+5d^{4}+3d^{2}-2)$\\
$\mathcal D_{3;5}$ & $48(5d^{4}-7d^{2}+2)$ & $\mathcal D_{23;23}$ & $-20d(d^{8}-71d^{6}-89d^{4}-86d^{2}+56)$\\
$\mathcal D_{4;4}$ & $-60(5d^{4}-7d^{2}+2)$ & $\mathcal D_{23;25}$ & $-12d(d^{6}+5d^{4}+3d^{2}-2)$\\
$\mathcal D_{01;01}$ & $48d(d^{14}-38d^{12}+39d^{10}+1919d^{8}-12374d^{6}+56940d^{4}-48552d^{2}+7168)$ & $\mathcal D_{23;34}$ & $-20d(d^{6}+5d^{4}+3d^{2}-2)$\\
$\mathcal D_{01;03}$ & $-48d(d^{12}+80d^{10}-191d^{8}-639d^{6}+25314d^{4}-21788d^{2}+2704)$ & $\mathcal D_{24;24}$ & $15d(d^{6}+5d^{4}+3d^{2}-2)$\\
$\mathcal D_{01;05}$ & $-48d(d^{10}-27d^{6}-127d^{4}+102d^{2}-12)$ & $\mathcal D_{012;012}$ & $d^{2}(d^{10}-6338d^{8}+8708d^{6}-133528d^{4}-82912d^{2}+76288)$\\
$\mathcal D_{01;12}$ & $-24d(36d^{12}+380d^{10}+527d^{8}-5232d^{6}+86092d^{4}-61496d^{2}+672)$ & $\mathcal D_{012;014}$ & $6d^{2}(d^{8}+242d^{6}+3292d^{4}+3792d^{2}-2224)$\\
$\mathcal D_{01;14}$ & $24d(16d^{10}+52d^{8}+555d^{6}+4072d^{4}-2228d^{2}-136)$ & $\mathcal D_{012;023}$ & $12d^{2}(d^{8}+242d^{6}+3292d^{4}+3792d^{2}-2224)$\\
$\mathcal D_{01;23}$ & $12d(22d^{10}+420d^{8}-4583d^{6}-31750d^{4}+22824d^{2}-352)$ & $\mathcal D_{012;025}$ & $6d^{2}(d^{6}-90d^{4}-188d^{2}+88)$\\
$\mathcal D_{01;25}$ & $-12d(22d^{8}-44d^{6}-255d^{4}+270d^{2}-56)$ & $\mathcal D_{012;034}$ & $-10d^{2}(d^{6}-90d^{4}-188d^{2}+88)$\\
$\mathcal D_{01;34}$ & $60d(2d^{8}-92d^{6}-517d^{4}+234d^{2}+16)$ & $\mathcal D_{012;045}$ & $6d^{2}(d^{4}+10d^{2}-4)$\\
$\mathcal D_{01;45}$ & $12d(2d^{6}-20d^{4}+13d^{2}-2)$ & $\mathcal D_{012;124}$ & $30d^{2}(d^{6}-90d^{4}-188d^{2}+88)$\\
$\mathcal D_{02;02}$ & $-60d(d^{12}-106d^{10}+171d^{8}+815d^{6}-27950d^{4}+22560d^{2}-1728)$ & $\mathcal D_{012;135}$ & $24d^{2}(d^{4}+10d^{2}-4)$\\
$\mathcal D_{02;04}$ & $120d(d^{10}+18d^{8}-141d^{6}-1093d^{4}+870d^{2}-96)$ & $\mathcal D_{012;234}$ & $60d^{2}(d^{4}+10d^{2}-4)$\\
$\mathcal D_{02;13}$ & $-48d(26d^{10}+52d^{8}+285d^{6}+2802d^{4}-1208d^{2}-256)$ & $\mathcal D_{013;013}$ & $-8d^{2}(d^{8}+242d^{6}+3292d^{4}+3792d^{2}-2224)$\\
$\mathcal D_{02;15}$ & $72d(12d^{6}+d^{4}+16d^{2}-8)$ & $\mathcal D_{013;015}$ & $-4d^{2}(d^{6}-90d^{4}-188d^{2}+88)$\\
$\mathcal D_{02;24}$ & $120d(7d^{8}+52d^{6}+165d^{4}+9d^{2}-44)$ & $\mathcal D_{013;024}$ & $-10d^{2}(d^{6}-90d^{4}-188d^{2}+88)$\\
$\mathcal D_{02;35}$ & $-72d(2d^{6}+4d^{4}+3d^{2}-2)$ & $\mathcal D_{013;035}$ & $-8d^{2}(d^{4}+10d^{2}-4)$\\
$\mathcal D_{03;03}$ & $-960d(3d^{8}-19d^{6}-161d^{4}+128d^{2}-14)$ & $\mathcal D_{013;123}$ & $-40d^{2}(d^{6}-90d^{4}-188d^{2}+88)$\\
$\mathcal D_{03;05}$ & $48d(d^{8}-2d^{6}-53d^{4}+39d^{2}-6)$ & $\mathcal D_{013;125}$ & $-24d^{2}(d^{4}+10d^{2}-4)$\\
$\mathcal D_{03;14}$ & $-120d(12d^{6}+d^{4}+16d^{2}-8)$ & $\mathcal D_{013;134}$ & $-40d^{2}(d^{4}+10d^{2}-4)$\\
$\mathcal D_{03;23}$ & $-120d(2d^{8}+84d^{6}+305d^{4}-78d^{2}-40)$ & $\mathcal D_{014;014}$ & $5d^{2}(d^{6}-90d^{4}-188d^{2}+88)$\\
$\mathcal D_{03;25}$ & $96d(d^{6}+8d^{4}-d^{2}-1)$ & $\mathcal D_{014;034}$ & $10d^{2}(d^{4}+10d^{2}-4)$\\
$\mathcal D_{03;34}$ & $120d(2d^{6}+4d^{4}+3d^{2}-2)$ & $\mathcal D_{014;124}$ & $30d^{2}(d^{4}+10d^{2}-4)$\\
$\mathcal D_{04;04}$ & $-60d(d^{8}-2d^{6}-53d^{4}+39d^{2}-6)$ & $\mathcal D_{023;023}$ & $20d^{2}(d^{6}-90d^{4}-188d^{2}+88)$\\
$\mathcal D_{04;24}$ & $-120d(d^{6}+8d^{4}-d^{2}-1)$ & $\mathcal D_{023;025}$ & $12d^{2}(d^{4}+10d^{2}-4)$\\
$\mathcal D_{12;12}$ & $-d(d^{12}-4519d^{10}-2385d^{8}-40170d^{6}-480320d^{4}+229056d^{2}+58496)$ & $\mathcal D_{023;034}$ & $20d^{2}(d^{4}+10d^{2}-4)$\\
$\mathcal D_{12;14}$ & $-6d(d^{10}+117d^{8}+2379d^{6}+8814d^{4}-2456d^{2}-1232)$ & $\mathcal D_{024;024}$ & $-15d^{2}(d^{4}+10d^{2}-4)$\\
$\mathcal D_{12;23}$ & $-12d(d^{10}+337d^{8}+1219d^{6}+6204d^{4}-716d^{2}-1312)$ &  & \\
\bottomrule
\end{tabular}
}
\end{table*}

\end{document}